\documentclass{ws-procs9x6}

\begin{document}

\title{The taming of the $\alpha$-vacuum}

\author{HAEL COLLINS
\footnote{\uppercase{C}urrent address:  \uppercase{D}epartment of \uppercase{P}hysics, \uppercase{U}niversity of \uppercase{M}assachusetts, \uppercase{A}mherst \uppercase{MA}\ \  01003, hael@physics.umass.edu.}
\footnote{\uppercase{T}his work was done in collaboration with \uppercase{R}ich \uppercase{H}olman and with the support of the \uppercase{D}epartment of \uppercase{E}nergy  (\uppercase{DE}-\uppercase{FG}03-91-\uppercase{ER}40682).}}

\address{Department of Physics \\
Carnegie Mellon University \\ 
Pittsburgh, PA\ \  15213, USA\\ 
E-mail:  hcollins@andrew.cmu.edu}

\maketitle

\abstracts{
When examining a field theory in a general state, the propagator must be made consistent with that state and new counterterms are allowed if the state breaks some of the space-time's symmetries.  This talk describes the specific example of how the propagator for the $\alpha$-vacua of de Sitter space must become the Green's function for two antipodal sources to obtain a renormalizable theory.
}

In the standard treatment of quantum field theory, the basic quantity we evaluate is the scattering matrix element between an initial and a final state.  Well away from the interaction region, these states are assumed to be the eigenstates of the free theory.  The background geometry is taken to be flat since the gravitational curvature is completely negligible over the scales at which the interactions occur.  For high energy collider experiments, each of these approximations is well justified.  

In inflation, the observed distribution of the large scale structure of the universe has its ultimate origin in the quantum fluctuations of a scalar inflaton field during an early epoch of extremely rapid expansion.  The environment in which these fluctuations occur differs dramatically from the pristine setting usually assumed for field theory.  The background is not flat.  Generally the space-time contains horizons, so there may be no time-like Killing vector with which to associate a conserved Hamiltonian.  The initial state is therefore often chosen for its symmetry rather than as an energy eigenstate.

A more mysterious feature of inflation results from the very expansion which makes it such an elegant explanation of the observed flatness and homogeneity of the universe on large scales.  Scales are constantly being redshifted during inflation so that with a sufficient amount of expansion, the fluctuations which have the correct size to seed the temperature fluctuations in the cosmic microwave background could have had their origin at sub-Planckian lengths.\cite{brandenberger}

To address such problems requires a careful treatment of quantum field theory in curved backgrounds starting from a general state.  This talk shows how to describe the propagation of this initial state information for a particular non-thermal, highly symmetric set of states in a de Sitter background, the $\alpha$-vacuum states.\cite{alpha}

de Sitter space corresponds to the space-time with a constant positive curvature.  A simple picture for de Sitter space\cite{houches} is obtained by embedding it in one spatial dimension higher where it corresponds to the hyperboloid given by 
\begin{equation}
- (X^0)^2 + (X^1)^2 + (X^2)^2 + (X^3)^2 + (X^4)^2 = H^{-2} . 
\label{dSembed}
\end{equation}
Here $H$ is the Hubble constant which is related to the cosmological constant through $\Lambda = 6H^2$.  An important property of de Sitter space is the existence of antipodal pairs of points, $\{X,X_A\}$, related by reflecting through the origin of the embedding coordinate system, $X_A^i\equiv - X^i$.

\begin{figure}[ht]
\centerline{\epsfbox{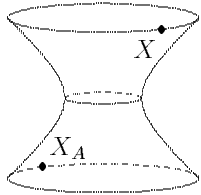}}
\caption{de Sitter space corresponds to the surface of a hyperboloid embedded in a flat space-time with one extra spatial dimension.  Antipodes are related by a reflection through the origin of the embedding space.
\label{hyper}}
\end{figure}

In terms of this embedding picture, the isometry group is easily visualized since it is the set of transformations that preserve the measure on the hyperboloid, $O(1,4)$.  The number of isometries of de Sitter space is thus exactly the same as in flat space.  Note that the subgroup of transformations that leave a particular point fixed in de Sitter space also leaves its antipode fixed.

Let us consider a scalar field $\Phi$ with a mass $m$.  In flat space the vacuum state for this field is the Poincar\' e-invariant, lowest energy eigenstate of the free Hamiltonian.  This Hamiltonian is globally conserved since one of the Killing vectors associated with the Poincar\' e symmetry is time-like everywhere.  In contrast, the horizons in de Sitter space prevent such a global time-like Killing vector for this background.  It is still possible to define a vacuum in the sense of being invariant under the full ten-dimensional $SO(1,4)$ symmetry group of de Sitter space.  This condition no longer selects a unique state but rather an infinite family of vacua.\cite{alpha}  These states are typically labeled by a complex parameter $\alpha$, with ${\rm Re}\, \alpha<0$, and are thus called the $\alpha$-vacua.

de Sitter space can be coordinatized conformally by defining 
\begin{eqnarray}
&X^0 = \tan t , 
\quad
X^1 = \sec t \cos\chi, 
\quad
X^2 = \sec t \sin\chi\cos\theta , &
\nonumber \\
&X^3 = \sec t \sin\chi\sin\theta\cos\phi , 
\quad\hbox{and}\quad
X^4 = \sec t \sin\chi\sin\theta\sin\phi . &
\label{conform}
\end{eqnarray}
These coordinates cover the entire space-time, with $t\in\{ -{\pi\over 2}, {\pi\over 2} \}$.  In terms of them, the metric becomes
\begin{equation}
ds^2 = \sec^2t \left[ dt^2 - d\Omega^2 \right] ,
\label{metric}
\end{equation}
with the metric on the three-sphere denoted by $d\Omega^2$.  The antipode of a point $x=(t,\chi,\omega,\phi)\equiv(t,\Omega)$ is $x_A=(-t,\Omega_A)$ where $\Omega_A$ are the coordinates the usual antipode on the three-sphere.

To define a vacuum state, we expand the field in a complete set of mode functions,
\begin{equation}
\Phi(x) 
= \sum_n \left[ \Phi_n(x)\, a_n + \Phi_n^*(x)\, a_n^\dagger \right] .  
\label{modes}
\end{equation}
The mode functions $\Phi_n(x)$ are the solutions of the second order Klein-Gordon equation.  One of the two constants of integration is fixed by the requirement that $\Phi(x)$ satisfies canonical commutation relations.  In flat space, the other constant would have been fixed by requiring the modes to be of positive energy.  Since this condition can no longer be imposed in de Sitter space, what is done instead is to require that the modes should become the standard flat space vacuum modes in the limit $H\to 0$ or equivalently at distances much shorter than the de Sitter radius, $H^{-1}$.  This requirement selects a unique state, the Bunch-Davies\cite{bunch} or Euclidean vacuum.  The modes and operators associated with this state will be written as $\Phi_n^E(x)$ and $a_n^E$ and the Bunch-Davies vacuum $|E\rangle$ satisfies, $a_n^E\, |E\rangle = 0$.

Since the short distance behavior of this state matches that of the flat space vacuum, the propagator $G_F^E(x,x')$ is still the Green's function associated with a single point source, 
\begin{equation}
\left[ \nabla_x^2 + m^2 \right] G_F^E(x,x') 
= {\delta^4(x-x')\over\sqrt{-g(x)}}
\label{pointsource}
\end{equation}
where $g$ is the determinant of the metric.  The solution to this equation is the usual time-ordered sum of two Wightman functions
\begin{eqnarray}
-i G_F^E(x,x') 
&=& \langle E | T \bigl( \Phi(x)\Phi(x') \bigr) | E\rangle 
\label{Eprop} \\
&=& \Theta(t-t')\, \langle E | \Phi(x)\Phi(x') | E\rangle 
+ \Theta(t'-t)\, \langle E | \Phi(x')\Phi(x) | E\rangle . 
\nonumber 
\end{eqnarray}
The renormalization of an interacting field theory in this vacuum proceeds exactly as in flat space---any divergences that occur can be absorbed by rescaling the field, mass and couplings or equivalently by adding local counterterms---since at the very short distances at which these divergences occur, $\ll H^{-1}$, the curvature of de Sitter space is unimportant and the modes were chosen to match the flat space modes.

Despite these appealing properties of the Bunch-Davies vacuum, the $\alpha$-vacua are also invariant under full $SO(1,4)$ de Sitter symmetry group.\footnote{Actually, the Bunch-Davies vacuum is $O(1,4)$ symmetric.   The breaking of $O(1,4)\to SO(1,4)$ by the $\alpha$-states allows extra counterterms during renormalization.}  Having defined the Bunch-Davies vacuum, the $\alpha$-vacua are readily defined as a Bogoliubov transform of it, 
\begin{equation}
a_n^\alpha = N_\alpha \bigl[ a_n^E - e^{\alpha^*} a_n^{E\dagger} \bigr] , 
\qquad
N_\alpha = \bigl( 1 - e^{\alpha+\alpha^*} \bigr)^{-1/2} . 
\label{alphadef}
\end{equation}
The $\alpha$-vacuum is then the state annihilated by this operator, $a_n^\alpha\, |\alpha\rangle = 0$.

While it may appear natural to generalize the propagator in this vacuum to be the same as that given by Eq.~(\ref{Eprop}), replacing the $|E\rangle$ state with the $|\alpha\rangle$ state, this construction misses an important point.  While the Wightman functions, $\langle\alpha | \Phi(x)\Phi(x') | \alpha\rangle$, are consistent with the new state, the time-ordering is not.  An interacting theory based on such a propagator manifests this inconsistency by no longer being renormalizable through the addition of a local set of counterterms\cite{fate,banks} or by the appearance of pinched singularities\cite{einhorn1}. 

To avoid these inconsistencies, let us define a new time ordering, 
\begin{eqnarray}
T_{\alpha} \bigl( \Phi(x) \Phi(x') \bigr)
&=& \Theta_\alpha(t,t')\, \Phi(x) \Phi(x') 
+ [\Theta_\alpha(t',t)]^*\, \Phi(x') \Phi(x)
\nonumber \\
&&
+\, \Theta^A_\alpha(t_A,t')\, \Phi(x_A) \Phi(x') 
+ [\Theta_\alpha^A(t',t_A)]^*\, \Phi(x') \Phi(x_A) ,
\qquad
\label{Talphadef}
\end{eqnarray}
where $t_A=-t$ and 
\begin{eqnarray}
\Theta_\alpha(t,t') 
&\equiv& \bigl[ 1 - e^{2\alpha} \bigr]^{-1} \bigl[ 
A_\alpha 
\left[ \Theta(t-t') + e^{2\alpha} \Theta(t'-t) \right]
- B_\alpha e^\alpha \bigr] 
\nonumber \\
\Theta_\alpha^A(t_A,t') 
&\equiv& \bigl[ 1 - e^{2\alpha} \bigr]^{-1} \bigl[ 
B_\alpha
\left[ \Theta(t_A-t') + e^{2\alpha} \Theta(t'-t_A) \right] 
- A_\alpha e^\alpha \bigr] . 
\label{ThetaAdefs}
\end{eqnarray}
The weighting factors, $A_\alpha$ and $B_\alpha$, are assumed to be real.  

Evaluating a $T_\alpha$-ordered product in an $\alpha$-vacuum yields a consistent $\alpha$ propagator, 
\begin{equation}
G_F^\alpha(x,x') 
= i \langle\alpha | T_{\alpha} \bigl( \Phi(x) \Phi(x') \bigr) | \alpha\rangle
= A_\alpha G_F^E(x,x') + B_\alpha G_F^E(x_A,x') . 
\label{alphaprop}
\end{equation}
Equation (\ref{pointsource}) reveals that the $\alpha$ propagator corresponds to the Green's function associated with {\it two\/} point sources located at antipodal points, 
\begin{equation}
\bigl[ \nabla^2_x + m^2 \bigr] G_F^\alpha(x,x') 
= A_\alpha {\delta^4(x-x')\over\sqrt{-g(x)}} 
+ B_\alpha {\delta^4(x_A-x')\over\sqrt{-g(x_A)}} . 
\label{twosource}
\end{equation}
This propagator forms the basis of a renormalizable perturbation theory for an interacting scalar theory in a classical gravitational background.\cite{einhorn2,lowe,taming}

What we have learned from the taming of the $\alpha$-vacuum is that perturbation theory constructed in a general state requires a corresponding modification of the propagator so that it is consistent with this state.  If the state breaks some of the background symmetry, new counterterms are also allowed that respect the remaining unbroken symmetry.  Generalizing this construction to an arbitrary expanding background provides a method for properly understanding the role of initial state effects during inflation.\cite{schalm,initprop}

\end{document}